\documentstyle[12pt,twoside]{article}

\def\be{\begin{equation}}
\def\ee{\end{equation}}
\def\bea{\begin{eqnarray}}
\def\ena{\end{eqnarray}}

\def\a{\alpha}
\def\b{\beta}

\def\d{\delta}
\def\e{\epsilon}           
\def\g{\gamma}
\def\h{\eta}

\def\j{\psi}

\def\m{\mu}
\def\n{\nu}
\def\o{\omega}
\def\p{\pi}                
\def\q{\theta}                    
\def\r{\rho}                      
\def\s{\sigma}                    
\def\t{\tau}

\def\del{\partial}

\def\cl{{\cal L}}

\catcode`@=11

\def\un#1{\relax\ifmmode\@@underline#1\else
$\@@underline{\hbox{#1}}$\relax\fi}

{\catcode`\'=\active \gdef'{{}^\bgroup\prim@s}}

\def\magstep#1{\ifcase#1 \@m\or 1200\or 1440\or 1728\or 2074\or 2488\or
       2986\fi\relax}   

\catcode`@=12

\def\bop#1{\setbox0=\hbox{$#1M$}\mkern1.5mu
	\vbox{\hrule height0pt depth.04\ht0
	\hbox{\vrule width.04\ht0 height.9\ht0 \kern.9\ht0
	\vrule width.04\ht0}\hrule height.04\ht0}\mkern1.5mu}

\def\slap#1#2{\setbox0=\hbox{$#1{#2}$}
        #2\kern-\wd0{\hbox to\wd0{\hfil$#1{/}$\hfil}}}

\def\leftrightarrowfill{$\mathsurround=0pt \mathord\leftarrow \mkern-6mu
       \cleaders\hbox{$\mkern-2mu \mathord- \mkern-2mu$}\hfill
       \mkern-6mu \mathord\rightarrow$}
\def\dvec#1{\vbox{\ialign{##\crcr
       \leftrightarrowfill\crcr\noalign{\kern-1pt\nointerlineskip}
       $\hfil\displaystyle{#1}\hfil$\crcr}}}          
\def\hook#1{{\vrule height#1pt width0.4pt depth0pt}}
\def\leftrighthookfill#1{$\mathsurround=0pt \mathord\hook#1
       \hrulefill\mathord\hook#1$}
\def\underhook#1{\vtop{\ialign{##\crcr                 
       $\hfil\displaystyle{#1}\hfil$\crcr
       \noalign{\kern-1pt\nointerlineskip\vskip2pt}
       \leftrighthookfill5\crcr}}}
\def\smallunderhook#1{\vtop{\ialign{##\crcr      
       $\hfil\scriptstyle{#1}\hfil$\crcr
       \noalign{\kern-1pt\nointerlineskip\vskip2pt}
       \leftrighthookfill3\crcr}}}

\def\sfrac#1#2{{\vphantom1\smash{\lower.5ex\hbox{\small$#1$}}\over
       \vphantom1\smash{\raise.4ex\hbox{\small$#2$}}}} 
\def\bfrac#1#2{{\vphantom1\smash{\lower.5ex\hbox{$#1$}}\over
       \vphantom1\smash{\raise.3ex\hbox{$#2$}}}}      
\def\afrac#1#2{{\vphantom1\smash{\lower.5ex\hbox{$#1$}}\over#2}}  
\def\on#1#2{{\buildrel{\mkern2.5mu#1\mkern-2.5mu}\over{#2}}}
\def\ddt#1{\on{\hbox{\LARGE .\kern-2pt.}}#1}             
\def\tdt#1{\on{\hbox{\LARGE .\kern-2pt.\kern-2pt.}}#1}   

\def\boxes#1{
       \newcount\num
       \num=1
       \newdimen\downsy
       \downsy=-1.5ex
       \mskip-2.8mu
       \bo
       \loop
       \ifnum\num<#1
       \llap{\raise\num\downsy\hbox{$\bo$}}
       \advance\num by1
       \repeat}
\def\boxup#1#2{\newcount\numup
       \numup=#1
       \advance\numup by-1
       \newdimen\upsy
       \upsy=.75ex
       \mskip2.8mu
       \raise\numup\upsy\hbox{$#2$}}

\newskip\humongous \humongous=0pt plus 1000pt minus 1000pt

\newif\ifdtup

\parskip=\medskipamount  
\def\baselinestretch{1.2}
\thicklines              
\def\border{
 }
\def\headpic{
 }
\def\title#1#2#3#4{\begin{document}
       \border
       \headpic
       {\hbox to\hsize{#4 \hfill ITP-SB-#3}}\par
       \begin{center}\vskip.8in minus.1in
       {\Large\bf #1}\\[.5in minus.2in]{#2}
       \vskip1.4in minus1.2in {\bf ABSTRACT}\\[.1in]\end{center}
       \begin{quotation}\par}
\def\author#1#2{#1\\[.1in]{\it #2}\\[.1in]}
\def\ITP{\footnote{Work supported by National Science Foundation
  grant PHY 89-08495.}\\[.1in] {\it Institute for Theoretical Physics\\
  State University of New York, Stony Brook, NY 11794-3840}\\[.1in]}
\def\endtitle{\par\end{quotation}\vskip3.5in minus2.3in\newpage}

\def\camera#1#2{
       \topmargin=.46in
       \textheight=22cm
       \textwidth=15cm
       \hsize=15cm
       \oddsidemargin=.28in
       \evensidemargin=.28in
       \marginparsep=0in
       \parindent=1.15cm
       \pagestyle{empty}
       \def\rm{\sf}
       \begin{document}
       \begin{center}{\Large\bf #1}\\[.5in minus.2in]{\bf #2}
       \vskip1in minus.8in {ABSTRACT}\\[.1in]\end{center}
       \renewcommand{\baselinestretch}{1}\small\normalsize
       \begin{quotation}\par}
\def\endabstract{\par\end{quotation}
       \renewcommand{\baselinestretch}{1.2}\small\normalsize}

\def\xpar{\par}                                       
\def\header{
 }
\def\letterhead{
 \header
 \font\sflarge=helvetica at 14pt   
 \leftskip=2.8in\noindent\phantom m\\[-.54in]
       {\large\sflarge STATE UNIVERSITY OF NEW YORK}
       {\scriptsize\sf INSTITUTE FOR THEORETICAL PHYSICS\\[-.07in]
       STONY BROOK, NY 11794-3840\\[-.07in]
       Tel: (516) 632-7979}
 \vskip.3in\leftskip=0in}
\def\letterneck#1#2{\par{\hbox to\hsize{\hfil {#1}\hskip 30pt}}\par
       \begin{flushleft}{#2}\end{flushleft}}
\def\letterhat{\parskip=\bigskipamount \def\baselinestretch{1}}
\def\head#1#2{\letterhat\begin{document}\letterhead\letterneck{#1}{#2}}
\def\multihead#1#2{\thispagestyle{empty}\setcounter{page}{1}
       \letterhead\letterneck{#1}{#2}}        
\def\shead#1#2{\letterhat\begin{document}\sletterhead
       \letterneck{#1}{#2}}
\def\multishead#1#2{\thispagestyle{empty}\setcounter{page}{1}
       \sletterhead\letterneck{#1}{#2}}
\def\multisig#1{\goodbreak\bigskip{\hbox to\hsize{\hfil Kind regards,
       \hskip 30pt}}\nobreak\vskip .5in\begin{quote}\raggedleft{#1}
       \end{quote}}
\def\sig#1{\multisig{#1}\end{document}}

\def\watch{
 \newcount\hrs
 \newcount\mins
 \newcount\merid
 \newcount\hrmins
 \newcount\hrmerid
 \hrs=\time
 \mins=\time
 \divide\hrs by 60
 \merid=\hrs
 \hrmins=\hrs
 \divide\merid by 12
 \hrmerid=\merid
 \multiply\hrmerid by 12
 \advance\hrs by -\hrmerid
 \ifnum\hrs=0\hrs=12\fi
 \multiply\hrmins by 60
 \advance\mins by -\hrmins
 \number\hrs:\ifnum\mins<10 {0}\fi\number\mins\space\ifnum\merid=0
 AM\else PM\fi}
\def\half{\frac{1}{2}}

\def\eq{\begin{equation}}
\def\eqe{\end{equation}}
\def\eqa{\begin{eqnarray}}
\def\eqae{\end{eqnarray}}

\def\be{\begin{equation}}
\def\ee{\end{equation}}
\def\bea{\begin{eqnarray}}
\def\ena{\end{eqnarray}}

\def\to{\rightarrow}

\def\sint{{\int
	   \hbox{\begin{picture}(4,2)(1.35,-1.1)
	   \put(0,0){\oval(4,2)}
	   \end{picture}}
	   \hskip-.28in\int}}

\def\dslash{\partial\!\!\! /  }
\begin{document}
\baselineskip=24pt
\begin{flushright}
{\sc ITP-SB}-96-65
\end{flushright}
{\Large{\bf
\centerline{ A continuous Wick rotation for spinor fields}

\centerline{and supersymmetry in Euclidean space}}}

\centerline{Peter van Nieuwenhuizen and Andrew
Waldron\footnote{Research supported by National
Science Foundation Grant PHY 930988. \\ e-mail:
vannieu@insti.physics.sunysb.edu\\ wally@insti.physics.sunysb.edu}}

\baselineskip=12pt

\centerline{Institute for Theoretical Physics}
\centerline{State University of New York at Stony Brook}
\centerline{Stony Brook, New York 11790-3840, USA}

\vspace{.20in}

\noindent {\bf Abstract:}  We obtain a continuous Wick rotation for
Dirac, Majorana and Weyl spinors $\j \to \exp ({1\over 2} \q \g^4
\g^5)\j$ which interpolates between Minkowski and Euclidean field
theories.

\baselineskip=16pt
\vspace{1.7cm}
In quantum field theory , the ``Wick rotation'' usually denotes
a rotation of $k_0$ in Greens functions. For bosonic fields,
such as the electromagnetic field, one may also define a Wick 
rotation on the fields themselves.
Recently, we have found a consistent way of defining a
Wick rotation on fermionic fields. The results will be published
in Physics Letters B (see also hep-th/9608174) where a complete
set of references can be found. Another article treating
the Wick rotation for fermionic fields in a canonical formalism 
is in preparation.

\section{Previous approaches.}
If one wants to study supersymmetric theories with instantons,
one obviously needs a Euclidean field theory for spinors.  Also for
the study of Donaldson invariants for compact Euclidean
manifolds , one must twist the bosonic fields of a $N=2$
supersymmetric model such that after twisting they become
half-integer spin fields.  A clear understanding of the relation
between Minkowski and Euclidean supersymmetry might help to
solve puzzles about reality issues. (Do twisting and Wick rotation
commute?)  Similarly, the recent interest in magnetic monopoles
and black holes leads one to ask questions about the Euclidean
formulation of such theories.

There are also non-supersymmetric reasons for being interested
in Euclidean field theories for spinors, and their relation to
Minkowski theories.  For example, in Fujikawa's approach to
anomalies, one regulates the Jacobian in the path integral with
operators such as $\rlap{\,/}D  \rlap{\,/}D$.  Only in Euclidean
space $\rlap{\,/}D$ is hermitean, and for that reason some authors
argue that this scheme only makes sense in Euclidean space.
(However, starting in Minkowski spacetime with nonhermitean
$\rlap{\,/}D\rlap{\,/}D$, one still can obtain momentum integrals
which become convergent after the usual analytic continuation in
$k_0$).  Further applications of Euclidean field theory are in
high-temperature field theory (where the transition from the real
time formalism to the imaginary time formalism involves a Wick
rotation), and in more formal aspects of field theory (for example
to justify in a more convincing way than just to add ``$i\e$" to
the time that Green's functions are vacuum expectation values, or
in the study of convergence of the perturbation series and the
effects of renormalons).

Despite these important fundamental issues and applications,
relatively little work has been done on the Wick rotation for
spinors.  If one looks at the lists of contents in recent text books
on quantum field theory at an advanced level, one finds at best a
discussion of the Wick rotation of the momentum variable $k_0$
in Green's functions, or a discussion of path integrals for bosons in
Euclidean space, but hardly any mention is made of the role
fermions must play in the Wick rotation.  Instead, most work has
concentrated on directly constructing a field theory in Euclidean
space whose Green's functions reproduce the analytically
continued Green's functions of the Minkowski theory.  Here (at
least) four approaches have emerged:
\begin{itemize}
\item [(i)] The Schwinger approach in which one constructs a
{\bf hermitean} action.  This precludes $N=1$ supersymmetry in
$d=4$ Euclidean space because there are no Majorana
spinors in Euclidean 
space.\footnote{A Majorana spinor is a spinor whose Majorana
conjugate $\j^T C$ is proportional to its Dirac conjugate
$\bar{\j}_D$.  In Minkowski spacetime, $\bar{\j}_D=\j^\dagger i \g^0$
but in Euclidean space $\bar{\j}_D = \j^\dagger$ (note that
the definition $\bar{\j}_D = -\j^\dagger\gamma^5$ in Euclidean
space is also
incompatible with the Majorana condition).}  For the $N=2$
model with Dirac spinors, Zumino constructed a hermitean
supersymmetric action, but one of the scalar fields has the wrong
sign in front of its kinetic term.  Some of the compact symmetries
become noncompact in the Euclidean theory, and vice-versa.
\item [(ii)] the Osterwalder-Schrader (OS) approach.  Here
hermiticity in Euclidean space is abandoned (which is not a
problem since hermiticity is primarily needed for unitarity, and
unitarity only makes sense in a theory with time).  The extension
of the OS formalism to Majorana spinors and 
supersymmetry was made by Nicolai.  The
basic idea of the OS approach for Dirac spinors is to view
$\bar{\j}$ and $\j$ in Euclidean space as two {\bf independent}
complex spinors, while for Majorana spinors in a $N=1$
Minkowski theory, one writes the action in Euclidean space as
$(\j^T C)\dslash \j$, just as in the Minkowski case, but one
considers $\j$ as complex (so no longer subject to a reality
constraint).  Since for susy one only needs properties of spinor
bilinears under transposition or the Fierz recoupling formula for
spinors, but not the reality condition on spinors, $N=1$ susy in
Euclidean space is obtained (but hermiticity is lost).
\item [(iii)] the Fubini-Hanson-Jackiw approach, in which one
views the radius in Euclidean  space as the time coordinate
(``radial quantization"). (Note that in string theory the map from
the cylinder to the plane $\exp (i\s + i t)\to \exp (i\s+\t)=z$ does
not correspond to radial quantization.  Rather, it is $\exp \t$
which plays the role of radius).
\end{itemize}
\noindent By far the most used approach is that of OS (and its
supersymmetric extension by Nicolai).  However, all these
approaches are either formulated in Minkowski spacetime or in
Euclidean space, but no continuous interpolation existed until the
work of Mehta.
\begin{itemize}
\item [(iv)] in the approach of Mehta, an external metric
$g_{\m\n}(\q)$  is introduced, which depends on an angle $\q \; (0
\leq \q \leq \p/2)$, such that for $\q=0$ one finds a Minkowski
theory for spinors, while at $\q = \p/2$ one obtains a Euclidean
theory.  He also rotates the Dirac matrices appearing in 
$\dslash$ (by a nonunitary
transformation) such that $\g^0$ 
and $\g^5$ become interchanged.
The $\g^4$ (where $\g^4 \equiv i \g^0$) in $\j^\dagger i\g^0$
is then reinterpreted as $\g^5_E$.
An appropriate choice of the matrix $g_{\m\n} (\q)$ 
yields the hermitean and
$SO(4)$ invariant action $\j^\dagger \g^5_E
(\g^4_E\del_4+\vec{\g}_E\cdot\vec{\del}+ m)\j$.
\end{itemize}
\noindent We shall further comment on some of 
these approaches in the final
section, but first we present our approach.

\section{The new Wick rotation.}
We shall construct a continuous Wick rotation depending on an angle
$\q \; (0 \leq \q \leq \p/2)$ for the basic fermionic {\bf fields}.  No
extraneous metric is introduced, and we do not rotate Dirac matrices
or any other constants; rather, as a result of rotating the spinor
fields, one can flip the rotation matrices from acting on the spinors to
acting on the Dirac matrices, and this will {\bf induce} a unitary
rotation of the Dirac matrices.  In fact, this rotation will turn out to
be a 5-dimensional Lorentz rotation, in the plane spanned by the
Minkowski and Euclidean time coordinates.  The basic problem is to
determine how the spinor fields transform under a Wick rotation;
once that problem is solved, all results should follow and no further
input should be necessary.

Of course, the Minkowski time coordinate $t$ rotates into the
Euclidean ``time" coordinate $\t$ according to
 \eq
t \to e^{-i\q} t_\q \; ; \; t_{\q=0} = t , t_{\q=\p/2} = \t
\label{wick1}
 \eqe
At all $\q$, $t_\q$ is real, hence the arrow indicates a
{\bf substitution} , not an equality.

In addition to transforming the time coordinate, we know already
from the example of the electromagnetic field $A_\m $ that one
should also transform fields.  Namely, the time component $A_0$
transforms in a way similar to $t$, namely\footnote{Fields transform
contragrediently to coordinates, so $A^0 \to e^{i\q} A^0_\q$.
Then $A_0 \to A_0^\q e^{-i\q}$ in order that $A^0$ and $A_0$ lead to
the same $A_4^E = A_E^4$.}
 \eq
A_0 (\vec{x}, t) \to e^{-i\q} A_0^\q (\vec{x}, t_\q)
\label{wick2}
 \eqe
At $t=0, A_0^{\q=0} (\vec{x}, t)$ equals the Minkowski field $A_0
(\vec{x}, t)$ while at $\q=\p/2, A_0^{\q=\p/2} (x,\t)$ is equal to the
fourth component $A_4^E (\vec{x}, \t)$ of the Euclidean field.  The
extra factor $e^{-i\q}$ becomes a factor $-i$ at $\q=\p/2$, such that
 \eq
\h^{\m\n} A_\m (\vec{x}, t) A_\n (\vec{x},t) \to \d^{\m\n}
A_\m ^E
(\vec{x}, \t) A_\n^E (\vec{x}, \t)
\label{wick3}
 \eqe
All this is well-known.

The main idea on which our new approach is based, is the
observation that the factor
$-i$ in $A_0 \to -iA_4^E$ is part of a {\bf matrix} $A_\m \to
S_\m{}^\n A_\n^E$ which happens to be diagonal for the
electromagnetic field
\eqa
&& A_\m (\vec{x}, t) \to M_\m{}^\n (\q) A_\n^\q (\vec{x}, t_\q)
\nonumber\\
&& M_\m{}^\n (\q) = \left( \begin{array}{llll} 1 \\ & 1 \\ & & 1 \\
 & & & e^{-i \q} \end{array} \right)
\label{wick4}
 \eqae
Thus, $A_\m$ transforms under Wick rotation as in the theory of
induced representations:  with an orbital part (the transformation of
coordinates) and a spin part (the matrix which acts on the indices of
the field).  In this sense, the ``Wick transformation" is not different
from Lorentz transformations, and as we shall see, there are more
analogies.

We therefore postulate that Dirac spinors transform under Wick
rotations as follows:
 \eq
\j ^\a (\vec{x}, t) \to S^\a{}_\b (\q) \j^\b_\q (\vec{x}, t_\q)
\label{wick5}
 \eqe
and we must now try to determine this matrix $S$.  However, before
going on with the spinorial case, we revert temporarily to the bosonic
case to deal with an unusual aspect having to do with complex
conjugation.  Suppose one were to Wick rotate a complex vector
boson field $W_\m$ (for example, the carrier of the weak
interactions).  Then one would require that
 \eq
\h^{\m\n} W_\m^* W_\n \to \d^{\m\n} (W^E_\m)^* W^E_\n
\label{wick6}
 \eqe
However, if $W_\m^*$ would transform with the complex conjugate
or hermitean conjugate matrix $M_\m{}^\n$, the phase factors
$(e^{i\q}$ and $e^{-i\q}$) would cancel each other.  To avoid this, we
must view complex conjugation as an antilinear operation for Wick
rotations (like time reversal) such that
\eq
W_\m^* (\vec{x}, t) \to M_\m{}^\n (\q) W_\n^* (\vec{x}, t)
\label{wick6a}
\eqe
or
\eq
W_\m^* (\vec{x}, t) \to M_\n{}^\m (\q) W_\n^* (\vec{x}, t)
\label{wick6b}
 \eqe
For bosons, (\ref{wick6a}) and (\ref{wick6b}) coincide because
$M_\m{}^\n$ is diagonal, but for fermions, only one of them will be
correct, as we shall discuss.

For $(\j^\a)^\dagger$ we consider two alternatives
 \eq
\j^\a (\vec{x}, t)^\dagger \to \j_\q^\b (\vec{x}, t_\q)^\dagger
S^\b{}_\a (\q)
\label{wick6c}
\eqe
or
\eq
 \j^\a (\vec{x}, t)^\dagger \to \j_\q^\b (\vec{x}, t_\q)^\dagger
 S^\a{}_\b
(\q)
\label{wick6d}
 \eqe
or in matrix notation $\j^\dagger \to \j^\dagger S$ or $\j^\dagger
\to \j^\dagger S^T$.  One might expect that only (\ref{wick6d}) is
correct but not (\ref{wick6c}) because the matrices $S$ should form a
representation of the Wick rotation:  $S(\q_1) S(\q_2) = S (\q_1 +
\q_2)$.  However, since the Wick rotation is an abelian group, $S(\q_1)$
and $S(\q_2)$ will commute so that both (\ref{wick6c}) and
(\ref{wick6d}) satisfy the group compositon law.  In fact,
(\ref{wick6c}) is the correct expression.

To fix the matrix $S$, we now impose three physical requirements.
Together they will determine $S$ uniquely.  We consider the laws in
(\ref{wick5}) and (\ref{wick6c}).  The alternative, (\ref{wick6d}), will be
shown later not to be viable.
\begin{itemize}
\item[(1)] The Wick rotation does not affect the space coordinates
$\vec{x}$, nor should it rotate the space components $\g^k$ of the
Dirac matrices.  Hence, $S(\q)$ should depend only on $\g^4$ and
$\g^5$ but not on $\g^1, \g^2$ or $\g^3$
\eq
[ S(\q) , \g^k ] = 0 \; {\rm for} \; k = 1,2,3
 \label{wick7}
 \eqe
We define $\g^5$ to be hermitean with square one $\g^5 \equiv \g^1
\g^2 \g^3 i \g^0$ and the Dirac matrices in Minkowski spacetime
satisfy    $\{ \g^\m, \g^\n \} = 2 \h^{\m\n}$.  For future use we
define $\g^4\equiv i\g^0$.
\item [(2)] In order that at $\q=\p/2$ the Euclidean action contains
the Euclidean Dirac operator
 \eq
\g_E^\m \del_\m = \g_E^k \del_k + \g_E^4 \del_4 \; ; \;
\{ \g_E^\m , \g_E^\n \} = 2 \d^{\m\n}
\label{wick9}
 \eqe
we require that $\j^\dagger i 
\g^0 \to \j^\dagger_\q S (\q) i \g^0 =
\j^\dagger_\q \; M S^{-1} (\q)$ for some matrix $M$.  Thus
 \eq
S ( \q) i \g^0 = M S (\q)^{-1}
\label{wick10}
\eqe
If this is the case, then the matrices
 \eqa
\g^k (\q) &\equiv& S^{-1} (\q) \g^k S (\q) = \g^k \nonumber\\
\g^4 (\q) &\equiv& S^{-1} (\q) \g^4 S (\q)
\label{wick11}
 \eqae
satisfy the Euclidean Clifford algebra at all $\q$.
  (At $\q = \p/2$ the
factor $e^{i\q}$ from $\del_t \to e^{i \q} \del_{t_\q}$ becomes a
factor $i$ which converts $\g^0$ into $i \g^0\equiv \g^4$.)
\item [(3)]  In order that at $\q=\p/2$ the action has a $SO(4)$
symmetry rather than a $SO(3,1)$ symmetry, we require that $M$
commutes with the $SO(4)$ generators $[\g_E^\m, \g_E^\n]$.
 \eq
[M, [ \g_E^\m , \g_E^\n ] ] = 0
\label{wick12}
 \eqe
\end{itemize}
\noindent {\bf The solution for the Wick rotation matrix
{\mbox{\boldmath $S(\q)$}} is}
{\mbox{\boldmath $ S (\q) = e^{{1\over 2} \g^4 \g^5 \q}$}}
where
 \eq
(\g^4)^2 = 1, (\g^5)^2 = 1 , (\g^4)^\dagger = \g^4,
(\g^5)^\dagger = \g^5.
\label{wick13}
 \eqe
It is clearly independent of $\g^k$ and it is unitary and satisfies
 \eq
S (\q) i \g^0  = i \g^0 S (\q)^{-1}
\label{wick14}
 \eqe
so $M = i \g^0$.  Explicit evaluation of $\g^4(\q)$ reveals
 \eq
\g^4 (\q) = S^{-1} (\q) \g^4 S (\q) = \g^4 \cos \q + \g^5  \sin \q
\label{wick15}
 \eqe
Hence, $\g_E^4 = \g^5$, and thus the Euclidean SO(4) generators
$[\g_E^\m , \g_E^\n]$ indeed commute with $M=i\g^0$.  In fact, it
is natural to define also a matrix $\g^5 (\q)$ by
 \eq
\g^5 (\q ) \equiv S^{-1} (\q) \g^5 S (\q)= - \g^4 \sin \q + \g^5 \cos \q
 \label{wick17}
 \eqe
so that $\g_E^5 = - \g^4 , \g_E^4 = \g^5$.
 The Euclidean action can now be written as
 \eq
\cl_E = \j_E^\dagger (\vec{x}, \t) \g_E^5 (\g_E^\m \del_\m + m )
\j_E (\vec{x}, \t)
\label{wick19}
 \eqe
and we see that it is hermitean and SO(4) invariant.  The Wick rotation
has not removed the matrix $i \g^0 $ ; rather it now appears as
$\g_E^5$.  Of course $\g_E^5 = \g_E^1 \g_E^2 \g_E^3
\g_E^4$.

One may ask whether this solution is unique.  In order that the Wick
rotation forms a one-parameter abelian group, one should be able to
write $S(\q)$ as $\exp \q N$ for some matrix $N$ which only depends
on $\g^4, \g^5$ and the unit matrix.  Pulling $i \g^0$ from the right to
the left of $S(\q)$ should result in $i \g^0 \exp (-\q N)$, and this fixes
$N$ to be a linear combination of $\g^4 \g^5$ and $\g^5$.
\eq
S (\q) = e^{\q (\a \g^4 \g^5 + \b \g^5)}
\label{wick20}
 \eqe
Finally, SO(4) invariance requires that $i \g^0$ anticommutes with
$S^{-1} (\q = \p/2) \g^4 S (\q = \p/2)$ (recall that $\g^k$ were
inert), which leads to the condition
that
 \eq
S (\q = \p/2)^2 + S (\q = \p/2)^{-2} = 0.
\label{wick21}
 \eqe
whose solution is $S(\q)$.

The alternative transformation law for $\j^\dagger$, namely
$\j^\dagger \to \j^\dagger S^T$, must be rejected, because at
$\q=\p/2$ the matrices $S(\q=\p/2)\equiv S$ must satisfy $S^T \g^4 =
(\a I + \b \g_E^5)S^{-1}$ in order that the Dirac operator be
obtained and SO(4) symmetry holds.  Using $\g_E^5 = S^{-1} \g^5 S$
leads to $SS^T = \a \g^4 + \b \g^5 \g^4$ which is inconsistent since in
general $\a \g^4 + \b \g^5 \g^4$ is not a symmetric matrix.

\section{Supersymmetry.}
Given a supersymmetric theory in Minkowski spacetime with Dirac
spinors $\j$, vector fields $V_\m$, scalars $A$ and pseudoscalars $B$,
we can obtain a corresponding supersymmetric theory in Euclidean
space by the following substitutions
\eqa
 && \j \to e^{\g^4 \g^5/4} \j_E , \j^\dagger \to \j^\dagger_E e^{\g^4
\g^5/4} \nonumber\\
&& V_\m = (V_0 , V_j) \to (- i V_4^E, V_j^E) ; d^4 x \to - i d^4 x_E
\nonumber\\
 &&  A \to A_E , B \to i B_E
\label{wick21a}
\eqae
The susy parameters are Dirac spinors and must be rotated
accordingly:
 \eq
\e \to e^{\g^4 \g^5/4} \e_E , \e^\dagger \to 
\e_E^\dagger e^{\g^4\g^{5/4}}
\label{wick22}
 \eqe
There is a consistency check on our procedure.  Consider $\d \j = M
\e$ for some matrix $M$.  To obtain $\d \j^\dagger$ one can either
first construct $\d \j_E$ and then take the hermitean conjugate, or
first take the hermitean conjugate and then perform the Wick
rotation.
\eqa
\d \j &=& M \e \Rightarrow S \d \j_E = (M_E) S \e_E \Rightarrow \d
\j_E^\dagger = \e_E^\dagger S^{-1} (M_E)^\dagger S \nonumber\\
\d \j^\dagger &=& \e^\dagger M^\dagger \Rightarrow \d
\j_E^\dagger S = \e_E^\dagger S (M^\dagger)_E \Rightarrow \d
\j_E^\dagger = \e_E^\dagger S (M^\dagger)_E S^{-1}
\label{wick23}
 \eqae
Hence, the matrix $M$ in $\d \j = M \e$ must satisfy
 \eq
S^{-1} (M_E)^\dagger S = S (M^\dagger)_E S^{-1}
\label{wick24}
 \eqe
where $M_E$ and $(M^\dagger)_E$ are obtained from $M$ and
$M^\dagger$ by using the substitutions in (\ref{wick21a}).  In
particular,
$V_\m^E, A^E$ and $B^E$ are to be considered real fields if $V_\m, A$
and $B$ were real.  The $N=2$ model of Zumino satisfies this
consistency check.  The origin of the factor $i$ in $B \to i B_E$
becomes clear if one takes as a model for a pseudoscalar
 \eq
B=\e^{\m\n\r\s} \del_\m \varphi^1 \del_\n \varphi^2 \del_\r
\varphi^3 \del_\s \varphi^4
\label{wick25}
\eqe
where $\varphi^i$ are real scalars. Clearly, $B$ picks up a factor $i$
under the Wick rotation, since one of the derivatives is $\del / \del t$.

The action one obtains in this way contains a matrix $\g_E^5$ as we
have seen, but for massless fields its presence may be overlooked
because one might use $i\g_E^5 \g_E^\m$ as Dirac matrices. In this
way one recovers the model of Zumino.

\section{Majorana and Weyl spinors.}
For Majorana spinors, we relax the reality condition on spinors, like
Nicolai.  Applying our transformation to the fields $\j$ and $\j^T C$ in
the Minkowski action, we transform them as
 \eq
\j \to S \j , (\j^T C) \to \j^T S^T C = \j^T (S^T CS)S^{-1}
\label{wick26}
 \eqe
The matrix $C_E \equiv S^T CS$ has the same symmetry properties as
$C$, hence for free spinors we get essentially the same action as
Nicolai (with $C_E$ instead of $C$)
 \eq
 \cl_E = \j_E^T C_E (\g_E^\m \del_\m + m) \j_E
\label{wick27}
 \eqe
However, all interactions in Euclidean space automatically follow from
the Minkowski theory; one must just insert the matrices $S$ and work
out consequences.  We have constructed some $N=1$ examples
which will be published elsewhere.

For Weyl spinors $\j_L = {1\over 2} (1+ \g^5) \j$ and $\j_L^\dagger =
{1\over 2} \j^\dagger (1+ \g_5)$ in Minkowski spacetime, are related
by complex conjugation.  Of course, in $d=4$ Minkowski spacetime one
can rewrite a Majorana spinor as a Weyl spinor.  Hence, for
consistency we relax the relation between $\j_L$ and $\j_L^\dagger$
before we begin the Wick rotation.   For clarity we write
$\j^\dagger =
\chi^\dagger$.  Then we start from
 \eq
\cl ({\rm Weyl}) = \chi^\dagger \left({1+ \g_5 \over 2}\right)
\g^4 \dslash  \left({1+\g_5 \over 2}\right) \j
\label{wick28}
 \eqe
and we rotate as before: $\j \to S \j_E, \chi^\dagger \to
\chi_E^\dagger S$.  Then
 \eq
\j_L \to S {(1+ \g_E^5)\over 2} \j_E , \chi_L^\dagger \g^4 \to
\chi_E^\dagger \g^4 \left({1- \g_E^5 \over 2}\right) S^{-1}
\label{wick29}
 \eqe
Hence, using $\g^4=-\g_E^5$, we find
\eq
  \cl ({\rm Weyl}) \to  \chi_E^\dagger \left({1-\g_E^5 \over 2}\right)
\g^4  \dslash^E \left( {1+\g_E^5\over 2}\right) \j =
\chi_R^{E\dagger} \dslash^E \j_L^E
\label{wick30}
 \eqe
As expected, the $\bar{\chi}_L \dslash \j_L$ pairing has
become a $(\chi_R^E)^\dagger \dslash^E \j_L^E$ pairing, as
required by SO(4) invariance.
\section{Discussion.}
An interpretation of our results in a five-dimensional spacetime with
signature $(4, 1)$ is possible. The matrix $\g^4 \g^5$ in $S(\q)$ is a
Lorentz generator in this space.  Also in the canonical work of OS
where all fields commute for all points in space, there are 5
dimensional echoes:  one seems to be working at equal time
$t=0$ in a 4-dimensional space $(\vec{x},\t)$.   Additional evidence are the
normalization factors $(k^2 + m^2)^{-1/2}$ where $k^2 = \vec{k}^2 +
k_4^2$ which appear in the second-quantized fermion fields of OS.
They are clearly the d=4 generalization of the usual d=3 factors
$(2\o)^{-1/2}$ where $\o^2 = \vec{k}^2 + m^2$. 
Further the mode operators in their work depend on Euclidean
four-momenta. Also in Schwinger's
work one finds 5-dimensional hints.  He introduces for each field $A(x)$
another field $B(x)$ satisfying $[A (x), B(y)] =  \d^4 (x-y)$, which can
again be viewed as equal-time 5 dimensional 
canonical commutation rules.

If one views the five-dimensional space as flat, with a flat vielbein
$e_A{}^M$, then the complex general coordinate transformation $t \to
e^{-i\q} t_\q$ acts on the index M of the vielbein, and one needs a
compensating Lorentz transformation to keep $e_A{}^M$ flat;  this
is just our 5-dimensional Lorentz rotation.  We are aware of many
loose ends which have to be better understood.

Coming back to the relation between Schwinger's approach and that
of OS, from our point of view, Schwinger's results are due to rotation
with our matrix S, (including the matrix $\g_E^5$), while OS results
can be viewed simply by defining $\j^\dagger \g_E^5$ as a new
complex spinor $\chi^\dagger$.  Of course, the OS approach is really
based on canonical quantization, and in Euclidean space one needs
twice as many creation and annihilation operators.  However, from a
path integral point of view, there is less difference.  Grassmann
integration over $\j$ and $\j^\dagger$ has as many degrees of
freedom as integration over $\chi^\dagger$ and $\j$.  The Grassmann
integral does not see reality properties, being based only on the rule
$\int d \j\j =1, \int d \j = 0$.  We believe that with our results in a
path integral context, the Wick rotation of spinors becomes as clear
as for bosons.

\end{document}